\documentclass[pra,twocolumn,showpacs]{revtex4}

\usepackage{amsfonts}
\usepackage{amsmath}
\usepackage{amssymb}
\usepackage{graphicx}
\usepackage{subfigure}
\usepackage{hyperref}

\newcommand{\pu}{^{\phantom{*}}}
\newcommand{\pd}{_{\phantom{n}}}
\newcommand{\efermi}{\varepsilon_{\rm F}\pu}
\newcommand{\kfermi}{k_{\rm F}\pu}
\newcommand{\lfermi}{l_{\rm F}\pu}
\newcommand{\gs}{g_{s}^{\phantom{*}}}
\newcommand{\vecr}{{\bf r}}
\newcommand{\vR}{{\bf R}}
\newcommand{\vk}{{\bf k}}

\begin{document}

\title{Density Correlations in Cold Atomic Gases: Atomic Speckles in the Presence of Disorder}

\author{Peter \surname{Henseler}}
\affiliation{Physikalisches Institut, Universit\"at Bonn, Nussallee 12, 53115 Bonn, Germany}
\author{Boris \surname{Shapiro}}
\affiliation{Department of Physics, Technion-Israel Institute of Technology, Haifa 32000, Israel}

\begin{abstract}
The phenomenon of random intensity patterns, for waves propagating in the presence of disorder, is well known in optics and in mesoscopic physics. We study this phenomenon for cold atomic gases expanding, by a diffusion process, in a weak random potential. We show that the density-density correlation function of the expanding gas is strongly affected by disorder and we estimate the typical size of a speckle spot, i.e., a region of enhanced or depleted density. Both a Fermi gas and a Bose-Einstein condensate (in a mean-field approach) are considered.
\end{abstract}


\maketitle

\section{Introduction}\label{section introduction}
A wave propagating in a random medium undergoes multiple
scattering from inhomogeneities. As a result, a complicated,
highly irregular, wave intensity pattern is formed. It is
characterized in statistical terms, with the help of various
correlation functions and probability distributions. Such
intensity patterns have been thoroughly studied for
electromagnetic waves in disordered dielectric media (optical
speckles) \cite{akk}. They are also well known for the Schr\"odinger
waves in disordered electronic conductors, where, due to the
electron scattering on impurities, a random distribution of
electron density and currents is established within the sample.
These "Schr\"odinger speckles" manifest themselves in various
mesoscopic effects, for instance, in the sample-to-sample
conductance fluctuations \cite{akk}.

Cold atomic gases offer a new playground for studying matter waves
in the presence of disorder. Unlike the case of electrons in
disordered conductors, for atomic gases one can directly measure
the atomic spatial density or, more precisely, density integrated
along the direction of imaging. There have been a number of recent
experiments \cite{lye, clement1, fort, shulte, clement2, yong}, as
well as some theoretical work \cite{san, paul} on the effect of
disorder on atomic Bose-Einstein condensates (BEC), in
quasi-one-dimensional geometry. Some of these
experiments reported observation of apparently random albeit
reproducible interference patterns - the atomic analog of the
Schr\"odinger wave speckles in disordered conductors.

In the present paper, we develop a theory for such "atomic speckles" in two and three spatial dimensions when, unlike the one-dimensional (1D) case, atoms can propagate by diffusion. The diffusion of the BEC in a random potential has been recently studied \cite{kuhn, sha, min} and it was pointed out that only some fraction of the condensate diffuses away, while the rest stays localized near its initial location (Anderson transition for BEC) \cite{sha, min}. Only the average density of the evolving atomic cloud was studied in Refs. \cite{sha, min}. The atomic speckles, however, are characterized by more complicated quantities, such as the density-density correlation function studied in the present paper. Before plunging into calculations, let us emphasize that we assume no (or negligibly small) disorder within the trap, i.e., the atoms are released from a "clean" trap and diffuse in the random potential. (A different setup is when the disorder, created within the trap, has a strong effect already at the equilibrium state of the system, causing a transition to an insulating phase \cite{shk}.) We focus mainly  on the case of Fermi gases (Section \ref{section fermi gas}). A brief discussion of the BEC case is given in Section \ref{section BEC}.

\section{Atomic speckles in a diffusing  Fermi gas}\label{section fermi gas}
We consider $N$ fermions, initially occupying the $N$ lowest energy eigenstates in a trap.
Assuming zero temperature and neglecting interactions we can write the quantum expectation value of
the particle density as
\begin{eqnarray}\label{trap density expectation value}
  \langle \hat{n}(\vecr) \rangle _0 & = & \gs\sum\limits_{n} f_{n}\pu \left| \phi_{n}\pu(\vecr) \right|^{2} ,
\end{eqnarray}
where $\gs=2S+1$ is the spin-degeneracy factor and $\phi_{n}\pu(\vecr)$ is the orbital part of an
 eigenstate in the trap. The factor $f_{n}\pu$ is unity for an occupied site and zero otherwise.

At time $t=0$ the Fermi gas is released from the trap and the
single-particle wave functions start to evolve according to the
Schr\"odinger equation
\begin{eqnarray}\label{Schroedinger equation}
  i\hbar\,\partial_{t}\pu \psi_{n}\pu(\vecr,t) & = & -\frac{\hbar^2}{2m} \triangle\psi_{n}\pu(\vecr,t) \, + \, V(\vecr)\, \psi_{n}\pu(\vecr,t) \: ,
\end{eqnarray}
where $V(\vecr)$ is the external static random potential. The quantum expectation value of the particle density, at time $t$, for a given realization of randomness is equal to
\begin{eqnarray}\label{density expectation value}
  \langle \hat{n}(\vecr,t) \rangle & = & \gs\sum\limits_{n} f_{n} \left| \psi_{n}\pu(\vecr,t) \right|^{2} \nonumber \\
  & = & \gs \int\! d^{d}R \int\! d^{d}R' \: G(\vecr,\vR,t) \, G^{*}(\vecr,\vR',t) \nonumber \\
  & & \times \sum\limits_{n} f_{n}\pu \, \phi_{n}\pu(\vR) \, \phi_{n}^{*}(\vR') \, , \quad
\end{eqnarray}
where $G(\vecr,\vR,t)$ is the retarded Green's function of the Hamiltonian
 in Eq.(\ref{Schroedinger equation}). The sum in Eq.(\ref{density expectation value}) can conveniently be written as
\begin{eqnarray}\label{sum trap states}
  \sum\limits_{n} f_{n}\pu \, \phi_{n}\pu(\vR) \phi_{n}^{*}(\vR') & = & \int\limits_{-\infty}^{\varepsilon_{\rm F}\pu}\! d\varepsilon \sum\limits_{n} \phi_{n}\pu(\vR) \phi_{n}^{*}(\vR') \, \delta(\varepsilon-\varepsilon_{n}\pu) \nonumber \\
  & = & -\frac{1}{\pi} \int\limits_{-\infty}^{\efermi}\! d\varepsilon \, {\rm Im} \mathcal{G}(\vR',\vR,\varepsilon) \, ,
\end{eqnarray}
where $\varepsilon_{n}\pu$ are the particle eigenenergies in the trap and $\varepsilon_{\rm F}\pu$
is the Fermi energy. The Green's function $\mathcal{G}({\it \vR',\vR},\varepsilon)$ refers to a particle in the trap and it should not be confused with the function $G(\vecr,\vR,t)$ which describes propagation from point $\vR$ to an observation point $\vecr$, upon the release of the gas from the trap. \\

The density $\langle \hat{n}(\vecr,t) \rangle$ is still a random
variable, in the sense that its value depends on the specific
realization of the random potential $V(\vecr)$. Denoting averaging
over realizations by an overbar, we write
\begin{eqnarray}\label{average density expectation value}
  \overline{\langle \hat{n}(\vecr,t) \rangle} & = & -\frac{\gs}{\pi}
   \int\! d^{d}R \int\! d^{d}R' \, \overline{G(\vecr,\vR,t) \, G^{*}\pd(\vecr,\vR',t)} \nonumber \\
   & & \quad \times \int\limits_{-\infty}^{\efermi}\!\! d\varepsilon \: {\rm Im} \mathcal{G}(\vR',\vR,\varepsilon) . 
\end{eqnarray}
The last integral in this expression contains information about
the initial state of the gas in the trap. The product of the two
propagators, $\overline{G\, G^{*}}$, propagates this information
in space and time.

In experiment, an ensemble of many random samples can be prepared
either by creating different realizations of the random potential
or by making slight changes in the Fermi energy $\efermi$. It is
well known in mesoscopic physics that, for a fixed realization of
randomness, various properties of a system are extremely sensitive
to the precise position of $\efermi$ \cite{akk}. Indeed, an
"ensemble of disordered conductors" is often produced from a
single sample, whose randomness is fixed by technology, by
changing the Fermi energy. In this respect, cold atomic gases
constitute a more convenient system for studying disorder related
effects. A different realization of disorder can be easily created
by changing the optical speckle pattern, on which the atoms are
being scattered. Furthermore, since the precise value of $\efermi$
cannot be controlled, from one experimental run to another, to an
accuracy higher than a few percent it appears that in experimenting
with cold atoms an "ensemble of random samples" is created
automatically, in a natural way.

The average product of the two Green's functions in Eq.(\ref{average density expectation value}) is a standard object in the theory of disordered systems. In the diffusion regime, considered in the present paper, it can be transformed as \cite{min}
\begin{eqnarray}\label{diffuson}
  & & \overline{G(\vecr,\vR,t) \, G^{*}\pd(\vecr,\vR',t)} \nonumber \\
  & & \qquad = \: -\frac{1}{\pi} \int\! d\varepsilon \,
  P_{\varepsilon}\pu(\vecr,\vR,t) \, {\rm Im}\bar{G}(\vR-\vR',\varepsilon)\, ,
\end{eqnarray}
where
\begin{eqnarray}\label{diffusion propagator}
  P_{\varepsilon}\pu(\vecr,\vR,t) & = & \frac{e^{-\left|\vecr-\vR\right|^2/4D_{\varepsilon}\pu t}}{(4\pi D_{\varepsilon}\pu t)^{d/2}}
\end{eqnarray}
is the diffusion propagator, in $d$ dimensions ($d=2,3$) and $D_{\varepsilon}\pu$ is the diffusion
coefficient for a particle at energy $\varepsilon$. The average Green's function is given by \cite{akk}
\begin{eqnarray}\label{average propagator}
  \bar{G}(\Delta R,\varepsilon) & = & G_{0}\pu(\Delta R, \varepsilon) \, e^{-\Delta R/2l_{\varepsilon}\pu} \, , \, \Delta R \: = \: \left| \vR-\vR' \right| , \quad
\end{eqnarray}
where $G_{0}\pu$ is the free Green's function and $l_{\varepsilon}\pu$ is the particle mean free path. The Eqs. (\ref{diffusion propagator}) and (\ref{average propagator}) for $P$ and $\bar{G}$ are valid only for sufficiently large values of the energy parameter $\varepsilon$, namely if $k_{\varepsilon}\pu l_{\varepsilon}\pu \gg 1$, where $k_{\varepsilon}\pu = \sqrt{2m\varepsilon/\hbar^2}$. Therefore, $\kfermi\lfermi$ is the essential parameter which determines the overall behavior of the gas, after switching off the trap, where $\kfermi$ and $\lfermi$ denote the value of the wave number and of the mean free path at the Fermi energy $\efermi$. Only for $\kfermi\lfermi\gg 1$, which is the case assumed in this paper, most of the atomic cloud will diffuse away from the trap. (A similar condition, with the chemical potential $\mu$ replacing the Fermi energy, is required for the diffusive behavior of a BEC cloud \cite{sha}.)

To facilitate analytic treatment, we assume that the size of the trap is much smaller than its distance from the observation point $\vecr$. Then, choosing the coordinate origin somewhere inside the trap, one can set $\vR=0$ in the second argument of the diffusion propagator, writing it simply as $P_{\varepsilon}\pu(r,t)$ ($r=|\vecr|$). Moreover, since upon release from the trap, each particle goes on its own (long) diffusive trajectory, the actual shape of the trap is of no importance. It is convenient to replace the actual harmonic trap by a cubic trap of size $L$, with periodic boundary conditions. Finally, Fourier transforming $\bar{G}$ in (\ref{diffuson}) with respect to $\vR-\vR'$, and performing integration over $\vR$ and $\vR'$ in (\ref{average density expectation value}), we obtain
\begin{eqnarray}\label{average density expectation value cubic trap}
  \overline{\langle \hat{n}(r,t) \rangle} & = & \frac{\gs L^{d}}{\pi^2} \int\! \frac{d^{d}k}{(2\pi)^d}
   \int\! d\varepsilon \, P_{\varepsilon}\pu(r,t) \, {\rm Im}\bar{G}(\vk,\varepsilon) \nonumber \\ 
  & & \qquad \times \int\limits_{-\infty}^{\efermi}\! d\varepsilon' \, {\rm Im}\mathcal{G} (\vk,\varepsilon') \, .
\end{eqnarray}
Since we assume no disorder in the trap, it follows that ${\rm Im}\mathcal{G}(\vk,\varepsilon') = -\pi\delta(\varepsilon'-\varepsilon_{\it k}\pu)$ and integration over $\varepsilon'$ results in a step function $\Theta(\kfermi - k)$. Furthermore, since $\kfermi\lfermi\gg 1$, it follows that the weak disorder condition, $k_{\varepsilon}\pu l_{\varepsilon}\pu \gg 1$, is satisfied for the great majority
of $k$'s in Eq. (\ref{average density expectation value cubic
trap}). Therefore, ${\rm Im}\bar{G}(\vk,\varepsilon)$ can be
approximated by $-\pi\delta(\varepsilon-\varepsilon_{\it k}\pu)$
and integration over $\varepsilon$ can be carried out, resulting
in
\begin{eqnarray}\label{average density expectation value final form}
  \overline{\langle \hat{n}(r,t) \rangle} & = & \gs L^{d} \int\limits_{k< \kfermi}\! \frac{d^{d}k}{(2\pi)^d} P_{k}\pu(r,t) \, ,
\end{eqnarray}
where the diffusion kernel $P_{k}\pu(r,t)$ is given by Eq. (\ref{diffusion propagator}),
with $R=0$ and $\varepsilon$ set equal to $\hbar^2\pd k^2\pd / 2m$. Equation (\ref{average density expectation value final form}) has a very simple interpretation: Particles, prior to their release from the trap, occupy all states up to $\kfermi$, with $L^{d}d^{d}k/(2\pi)^{d}$ being the number of particles in the element $d^{d}k$. When the trap is switched off, particles start diffusing and $P_{k}\pu(r,t)$ is the probability density that a particle with wave number $k$ will reach point $\vecr$ in time $t$. Integration over $\vk$ gives the average particle density $\overline{\langle \hat{n}(r,t) \rangle}$. An equation similar to (\ref{average density expectation value final form}) exists also for a BEC \cite{sha}. It was emphasized there, and the same is true for the present case of a Fermi gas as well, that some fraction of the released particles will not propagate by diffusion but will remain localized near their original location. For $\kfermi\lfermi\gg 1$ the fraction of such particles is small and we ignore them in the present work. \\

We turn now to the calculation of the density-density correlation function. Let us first note that,
already in the absence of disorder, a Fermi gas possesses some subtle density correlations of purely
quantum nature \cite{ll}. The expectation value, $\langle \hat{n}(\vecr, t) \rangle$, is ignorant
about these correlations. However, as has been particularly emphasized in \cite{leg,adl}, a single
imaging experiment (with sufficient resolution) does not measure $\langle \hat{n}(\vecr, t) \rangle$
but rather one particular event, i.e., some particular density pattern, $n(\vecr, t)$, whose
probability is dictated by the many-body wave function of the system.
 The density correlation function can be extracted from such noisy density patterns by averaging
  the product  $n(\vecr)n(\vecr')$ over many experimental
realizations, taken under (as far as possible) identical
experimental conditions, for fixed $\vecr, \vecr'$ and $t$. Such
averaging would yield  the theoretically calculated quantum
expectation value $\langle \hat{n}(\vecr)\hat{n}(\vecr') \rangle$.
It should be noted, however, that already a single experimental
image, although noisy and "grainy", contains information about the
density correlation function. This information will be revealed by
taking the product $n(\vecr)n(\vecr +\Delta\vecr)$ and averaging
it over many points $\vecr$, for fixed $\Delta\vecr$ (the
equivalence of the two averaging procedures constitutes the
"ergodic assumption" in mesoscopic physics \cite{akk}).

For fermions confined to a
trap of size $L$, the density-density correlation function
exhibits rapidly decaying oscillations with a  characteristic
period  $\Delta x_{0}\pu \approx k_{\rm F}^{-1} \simeq LN^{-1/d}$
\cite{ll, foot1}. When the gas is released from the trap it starts
expanding, and, for times $t>(L/v_{\rm F}\pu) \equiv t_{0}\pu$,
the size of the cloud grows linearly with time and so does the
correlation length $\Delta x(t) \approx \Delta x_{0}\pu \cdot
t/t_{0}\pu$ \cite{ng}. Thus, roughly speaking, the free, ballistic
expansion amplifies the scale of correlations by the factor
$t/t_{0}\pu$.

Below we show that in the presence of a random potential, i.e.,
when the expansion is diffusive instead of ballistic, the picture
is different: The size of the atomic cloud grows as $\sqrt{t}$
whereas the short-range correlations do not get amplified at all.
The density-density correlation function in the presence of
disorder is defined as
\begin{eqnarray}\label{definition density-density correlation function}
  C(\vecr,\vecr',t) & = & \overline{\langle \hat{n}(\vecr,t) \hat{n}(\vecr',t) \rangle} \, - \, \overline{\langle \hat{n}(\vecr,t)\rangle} \, \overline{\langle \hat{n}(\vecr',t)\rangle} \nonumber \\
  & & - \, \delta(\vecr-\vecr') \overline{\langle \hat{n}(\vecr,t)\rangle} \, . 
\end{eqnarray}
The last term describes trivial correlations, which exist already
in a classical ideal gas and which are commonly subtracted, in
order to isolate the nontrivial correlations \cite{ll}. There are
two kinds of averaging in (\ref{definition density-density
correlation function}): The quantum mechanical averaging, for a
given realization of disorder, and averaging over the ensemble of
different realizations. The first averaging is straightforward and
leads to \cite{ll}:
\begin{eqnarray}\label{quantum correlations}
  \langle \hat{n}(\vecr,t) \hat{n}(\vecr',t) \rangle & = & g^{2}_{s} \sum\limits_{n} f_{n}\pu \left|\psi_{n}\pu(\vecr,t)\right|^2  \sum\limits_{m} f_{m}\pu \left|\psi_{m}\pu(\vecr',t)\right|^2 \nonumber \\
   & & - \, \gs \left| \sum\limits_{n} f_{n}\pu \psi_{n}^{*}(\vecr,t) \psi_{n}\pu(\vecr',t) \right|^2 \nonumber \\
   & & + \, \gs \delta(\vecr-\vecr') \sum\limits_{n} f_{n}\pu \left|\psi_{n}\pu(\vecr,t)\right|^2 \, ,
\end{eqnarray}
where the absence of overbars indicates that this expression refers to a specific realization of the random potential. Next, we must average (\ref{quantum correlations}) over the disorder. This involves averaging products of four single-particle wave functions. For short-range correlations, i.e., on a scale smaller than the mean free path $l_{\varepsilon}\pu$, such averages decouple into products of pairwise averages \cite{akk}. For instance,
\begin{eqnarray}\label{decoupling approximation}
  & & \overline{\psi_{n}^{*}(\vecr,t)\psi_{n}\pu(\vecr',t)\psi_{m}^{*}(\vecr',t)\psi_{m}\pu(\vecr,t)} \nonumber  \\
  & & \approx \: \overline{\psi_{n}^{*}(\vecr,t)\psi_{n}\pu(\vecr',t)}\: \overline{\psi_{m}^{*}(\vecr',t)\psi_{m}\pu(\vecr,t)} \nonumber \\
  & &  \quad + \, \overline{\psi_{n}^{*}(\vecr,t)\psi_{m}\pu(\vecr,t)}\: \overline{\psi_{m}^{*}(\vecr',t)\psi_{n}\pu(\vecr',t)} \, , \quad \left|\vecr-\vecr'\right| < l  . \nonumber \\
\end{eqnarray}
Performing such decoupling in Eq. (\ref{quantum correlations}) and subtracting the product of averaged densities, $\overline{\langle n(\vecr,t)\rangle} \, \overline{\langle n(\vecr',t)\rangle}$, we obtain
\begin{eqnarray}\label{correlation function disorder + quantum}
  C(\vecr,\vecr',t) & = & \sum\limits_{n,m} f_{n}\pu f_{m}\pu \left\{ -\gs A_{nn}\pu(\vecr,\vecr',t) A_{mm}^{*}(\vecr,\vecr',t)\phantom{\sum} \right. \nonumber \\
  & & \left. - \, \gs A_{nm}\pu(\vecr,\vecr,t) A_{nm}^{*}(\vecr',\vecr',t) \right. \nonumber \\
  & & \left. + \, g^2_{s} \left| A_{nm}\pu(\vecr,\vecr',t) \right|^2\pd \right\} \, , 
\end{eqnarray}
where
\begin{eqnarray}\label{definition A_nm}
  A_{nm}\pu(\vecr,\vecr',t) & \equiv & \overline{\psi_{n}^{*}(\vecr,t)\psi_{m}\pu(\vecr',t)} \, .
\end{eqnarray}
The first term in (\ref{correlation function disorder + quantum}) describes quantum correlations, due to the Pauli exclusion principle. In particular, for $\vecr'\rightarrow \vecr$, it approaches the value $-\gs\overline{\langle \hat{n}(\vecr,t) \rangle}^{2}$ and, thus, it is proportional to $N^2$. The third term is of "classical" origin, in the sense that it originates from the interference between multiply scattered waves. It contributes positive correlations, similarly to speckle pattern in optics. However, in contrast to a single frequency laser speckle, here there are many waves with different frequencies. Since contributions from different frequencies should be added incoherently, i.e., intensities (rather than amplitudes) are summed up, the third term is proportional to $N$ and it will be neglected. The second term in (\ref{correlation function disorder + quantum}) is a combination of quantum and classical correlations. Its sign and the factor $\gs$ originate from the exclusion principle. Since, however, wave functions at different energies are essentially uncorrelated, this term is also proportional to $N$ and can be neglected in the large-$N$ limit. Thus, keeping only the first term in (\ref{correlation function disorder + quantum}), we obtain
\begin{eqnarray}\label{effective correlation function}
C(\vecr,\vecr',t) & = & -\gs \left| \sum\limits_{n} f_{n}\pu
A_{nn}\pu(\vecr,\vecr',t) \right|^{2} \nonumber \\
& \equiv & -\gs \left|
F(\vecr,\vecr',t) \right|^{2} \, .
\end{eqnarray}
The function $F(\vecr,\vecr',t) \equiv \sum\limits_{n} f_{n}\pu
\overline {\psi_{n}^{*}(\vecr,t)\psi_{n}\pu(\vecr',t)}$ can be
expressed in terms of the Green's functions, in complete analogy
with the earlier derivation for the average density. The only
difference is that now there are two "observation points", $\vecr$
and $\vecr'$. The resulting expression for $F(\vecr,\vecr',t)$ (compare
to Eq. (\ref{average density expectation value})) is
\begin{eqnarray}\label{correlation functions Green's functions}
  F(\vecr,\vecr',t) & = & - \frac{1}{\pi} \int\! d^{d}R \int\! d^{d}R' \, \overline{G(\vecr,\vR,t)\, G^{*}\pd(\vecr',\vR',t)} \nonumber \\
  & & \qquad \times \int\limits_{-\infty}^{\efermi}\! d\varepsilon \, {\rm Im}\mathcal{G}(\vR',\vR,\varepsilon) \, . \qquad
\end{eqnarray}
In order for the average product of Green's functions not to be exponentially small, the pair of points $\vecr,\vecr'$ should not be separated by more than the mean free path. On the other hand, $\left| \vecr-\vR \right|$ is much larger than $\lfermi$. The computation of the average product in (\ref{correlation functions Green's functions}) is straightforward and the result is
\begin{eqnarray}\label{diffusion approximation correlation function}
  \overline{G(\vecr,\vR,t)\, G^{*}\pd(\vecr',\vR',t)} \nonumber & = & -\frac{1}{\pi} \int\! d\varepsilon \, f(\Delta r,\varepsilon) \, P_{\varepsilon}\pu(\vecr,\vR,t) \nonumber \\
  & & \qquad\quad \times {\rm Im}\bar{G}(\vecr-\vR',\varepsilon) \, , \qquad
\end{eqnarray}
where the extra factor $f(\Delta r,\varepsilon)$, as compared to Eq. (\ref{diffuson}), is given by
\begin{eqnarray}\label{definition f-function}
  f(\Delta r, \varepsilon) & = & -\frac{1}{\pi\nu_{\varepsilon}\pu} {\rm Im}\bar{G}(\Delta r, \varepsilon) \, , \quad \Delta r \: = \: \left| \vecr-\vecr'\right| \, . \quad 
\end{eqnarray}
Following the same line of derivation as for the average density,
 Eq. (\ref{average density expectation value final form}), we arrive at
\begin{eqnarray}\label{correlation function final form}
  F(\vecr,\vecr',t) & = & L^{d}\pd \int\limits_{k<\kfermi} \frac{d^{d}k}{(2\pi)^{d}\pd}
   P_{k}\pu(r,t) f(\Delta r, \varepsilon_{\it k}) \, .
\end{eqnarray}
For $d=3$ \cite{akk},
\begin{eqnarray}\label{f-function 3d}
  f(\Delta r, \varepsilon_{\it k}) & = & \frac{\sin(k \Delta r)}{k \Delta r} e^{-\Delta
  r/2l_k}.
\end{eqnarray}
While the kernel $P_k$ and the mean free path $l_k$ are slowly
changing with $k$, the function $f(\Delta r, \varepsilon_{\it
k})$, for $\Delta r \gg k_F^{-1}$, contains  a rapidly oscillating
factor within the integration region in (\ref{correlation function
final form}). Taking the "slow" functions out of the integral and
computing the remaining integral we obtain
\begin{eqnarray}\label{density-density correlation function 3d result}
  C(\vecr,\vecr',t) & = & -\gs L^{2d}\pd P_{\kfermi}\pu(r,t)^{2}\pd e^{-\Delta r/\lfermi} \frac{k_{\rm F}^{6}}{4\pi^{4}\pd} \nonumber \\
  & & \times \frac{\left[\sin(\kfermi\Delta r) \, - \, \left( \kfermi\Delta r \right)
    \cos(\kfermi\Delta r)\right]^{2}\pd}{\left(\kfermi\Delta
    r\right)^{6}\pd}, \qquad
\end{eqnarray}
where $P_{k}\pu$ and $l_{k}\pu$ are taken at $k=k_F$. It is convenient to
normalize the correlation function by the average density which
yields
\begin{eqnarray}\label{normalized density-density correlation function}
  & & I(\vecr,\vecr',t) \: \equiv \: \frac{C(\vecr,\vecr',t)}{\overline{\langle \hat{n}(r,t) \rangle} \, \overline{\langle \hat{n}(r',t) \rangle}} \nonumber \\
  & = & -\frac{9}{\gs} e^{-\Delta r/\lfermi} \frac{\left[\sin(\kfermi\Delta r) \, - \, \left( \kfermi\Delta r \right) \cos(\kfermi\Delta r)\right]^{2}\pd}{\left(\kfermi\Delta r\right)^{6}\pd} \, . \qquad
\end{eqnarray}
The analogous calculation for $d=2$ yields
\begin{eqnarray}\label{normalized density-density correlation function 2d}
  I(\vecr,\vecr',t) & = & -\frac{4}{\gs} e^{-\Delta r/\lfermi} \frac{{\rm J}_{1}\pu(\kfermi\Delta r)^{2}\pd}{\left(\kfermi\Delta r\right)^{2}\pd} \, ,
\end{eqnarray}
where ${\rm J}_{1}\pu(\kfermi\Delta r)$ is the Bessel function of the first kind.

\begin{figure}
 \begin{center}
   \subfigure[]{\includegraphics[scale=0.65]{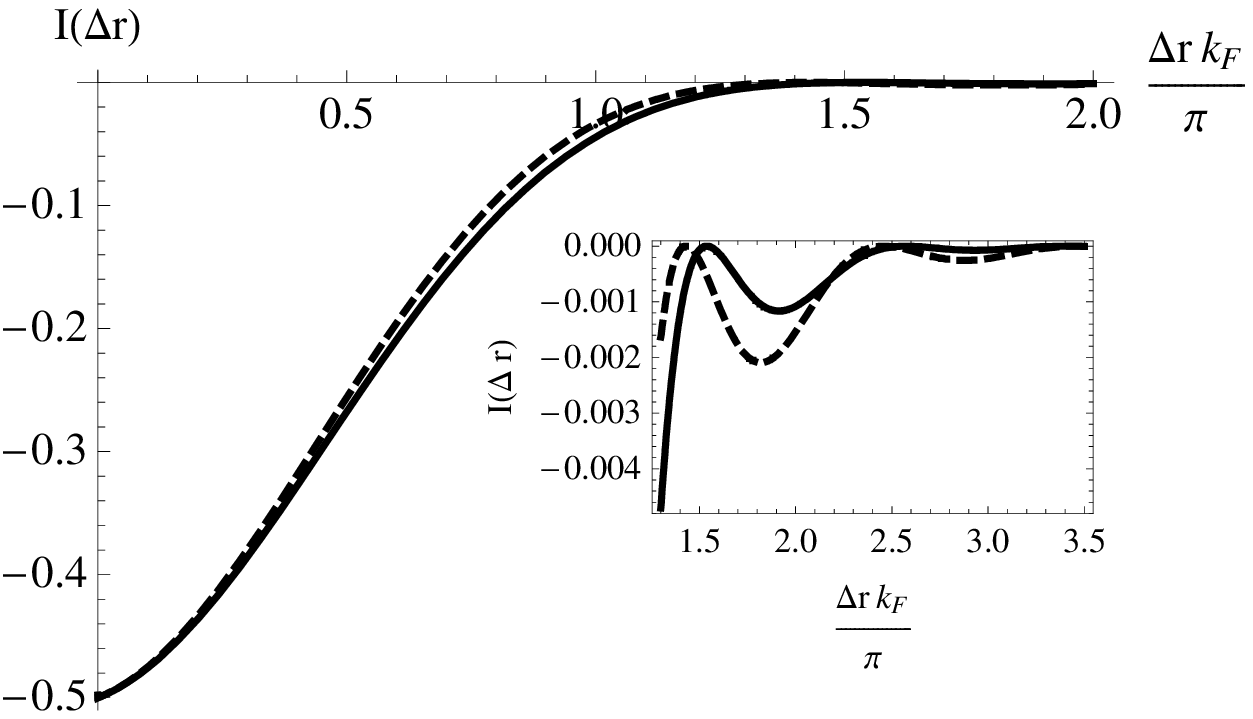} \label{fig correlation function 3d}}
   \hfil
   \subfigure[]{\includegraphics[scale=0.65]{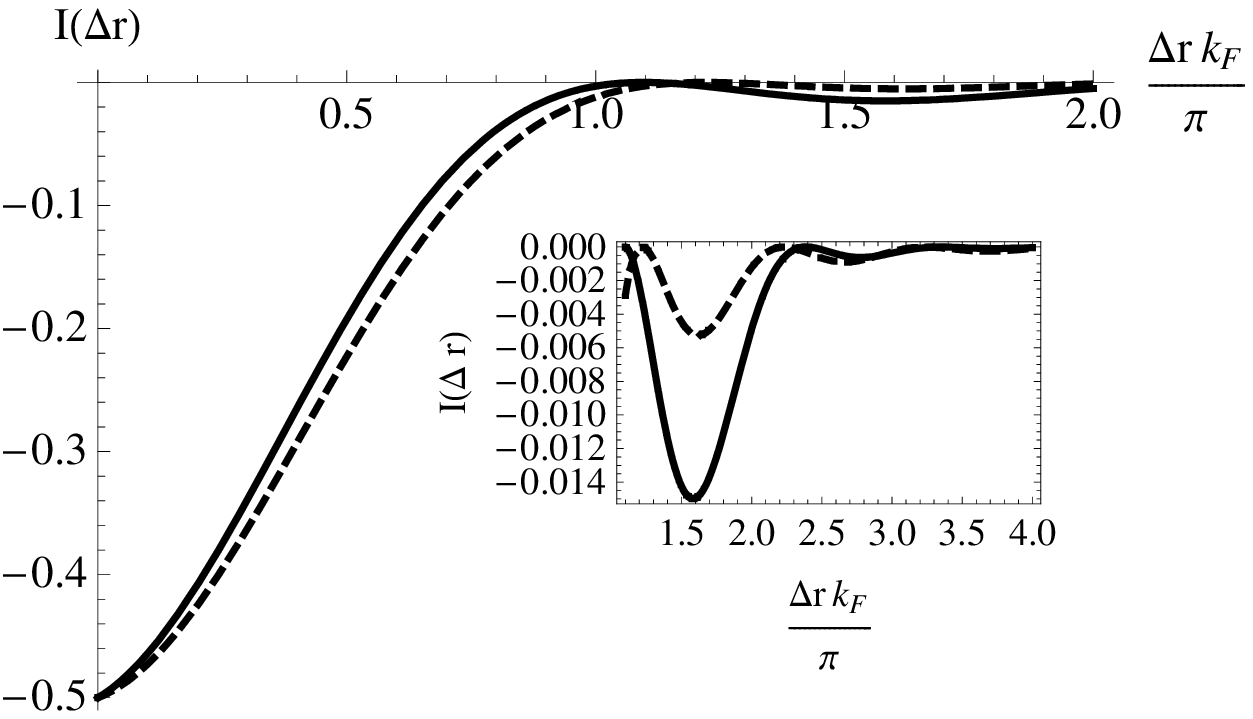} \label{fig correlation function 2d}}
   \caption{The normalized density-density correlation function $I(\Delta
r,t)$ of a Fermi gas with  $g_s=2$, for  $k_{\rm F} l_{\rm F} =
10$, $r=50 l_{\rm F}$ and $t=3r^{2}/(2dD_{k_{\rm F}})$. (a)
corresponds to $d=3$ and (b) to $d=2$. The dashed curves in (a)
and (b) correspond to  Eqs. (\ref{normalized density-density
correlation function}) and (\ref{normalized density-density
correlation function 2d}), respectively. The solid curves are the
result of an exact numerical integration, as described in the
text. The insets show the
oscillating decay of $I(\Delta r,t)$ with increasing
$\Delta r$.}
 \label{fig correlation function}
 \end{center}
\end{figure}

The normalized density-density correlation functions, given in
Eqs. (\ref{normalized density-density correlation function}) and
(\ref{normalized density-density correlation function 2d}) are
plotted in Fig. \ref{fig correlation function} (dashed lines).
The solid lines in this figure represent the corresponding
functions obtained by computing the integral in
Eq. (\ref{correlation function final form}) numerically, without
taking the "slow" functions out of the integral. In this
computation a white noise random potential has been used which
yields $l_{k}\pu=const$ in three dimensions and $l_{k}\pu\sim k$ in two dimensions (this
implies that the diffusion coefficient is proportional to $k$ and
$k^2$ in  three or two dimensions, respectively). Figure \ref{fig
correlation function} demonstrates that for the decaying envelope
of $I(\Delta r)$ the agreement between the numerically exact
results and the approximate expressions, Eqs. (\ref{normalized
density-density correlation function}) and (\ref{normalized
density-density correlation function 2d}), is quite good. For the
oscillations the agreement is only qualitative. Due to the rapid
decay of the envelope function, the oscillations are rather small
and clearly visible only on an amplified scale, as shown in the
insets to the figure.

It is quite remarkable that, while the atomic cloud keeps expanding, the local normalized density correlations, Eqs. (\ref{normalized density-density correlation function}) and (\ref{normalized density-density correlation function 2d}), do not depend on time. In particular, the characteristic length of
oscillations, $\Delta r \sim k_{\rm F}^{-1}$, remains the same as
for the gas in the trap, prior to its release. This behavior is in sharp contrast
with that for a free ballistic expansion, when the spatial
oscillation period is growing linearly in time, just as the
average interparticle distance. The essential difference between
ballistic and diffusive cases can be traced to the evolution of
the phase of a wave function. In the ballistic expansion the
characteristic spatial period of a wave packet keeps increasing.
(This statement can be easily verified by taking an initial wave
packet of a Gaussian shape, $\phi(r)\sim \exp(-\frac{r^2}{a^2} +
i\frac{r^2}{b^2})$, with $b\ll a$, whose time evolution is exactly
soluble.) On the contrary, in the diffusive expansion the phase of
a wave function gets randomized after few scattering events and no
long-range order in the phase can be established. An oscillating
wave function with some period $b$, subjected to diffusive
evolution, will locally look as a plane wave, with the same
period, where "locally"  means on a scale smaller than the mean
free path. Thus, at any time $t$, the diffusing wave function can be
viewed as made up of "patches" of plane waves, of size $\lfermi$
each, but with no phase relation among different patches. This
observation explains the somewhat counterintuitive behavior of the
correlations, namely, that the correlation functions in
Eqs. (\ref{normalized density-density correlation function}) and
(\ref{normalized density-density correlation function 2d}) remain
stationary, while the interparticle distance increases under the
expansion. This leads to an increase of the relative fluctuation
of the particle number, in a given volume. For a homogeneous Fermi
gas in equilibrium (in three dimensions) the particle number variance
$\overline{\Delta N^2}$, in a certain volume, is not equal to the
average number of particles $\overline N$ (in the same volume) but
is proportional to $\overline{N}^{2/3}\log\overline{N}$
\cite{ast,cast}. This means that, due to correlations, the Fermi
gas possesses some kind of "rigidity". (The effect is particularly
spectacular in 1D, where $\overline{\Delta N^2}$ grows only as
$\log\overline{N}$). Free expansion of the gas, in the absence of
disorder, does not affect this rigidity, because, as was already
mentioned, the scale of correlations is amplified in exact
proportion to the interparticle distance. The disorder disrupts
this proportionality and leads to destruction of rigidity and to
the $\overline{\Delta N^2}=\overline N$ behavior (in the long time
limit). Therefore, the image of a Fermi gas, expanding in the
presence of disorder, should look more "grainy" than the image of
a freely expanding gas. This effect might be observable
experimentally.

\section{Correlations in  a diffusing BEC: a mean-field approach}\label{section BEC}
So far we considered the dynamics of a degenerate Fermi gas. We
now briefly discuss the case of a BEC, expanding in the presence
of a random potential. Within the mean-field approach the BEC is
described by a macroscopic wave function, $\Psi({\bf r}, t)$,
whose dynamics satisfies the Gross-Pitaevskii equation
\begin{eqnarray}\label{G1}
 i\hbar\, \partial_{t}\pu \Psi(\vecr,t) & = &
-\frac{\hbar^2}{2m}\,\triangle\Psi(\vecr,t)
 \: + \: V({\bf r})\Psi(\vecr,t) \nonumber \\
 & & +\: g |\Psi(\vecr,t)|^2 \Psi(\vecr,t) \, ,
\end{eqnarray}
where $g$ is the interaction parameter related to the scattering
length (we assume positive $g$, i.e., repulsive interactions).
Equation (\ref{G1}) describes the evolution of the condensate, in the
random potential $V({\bf r})$, upon its release from the trap. We
assume an isotropic harmonic trap, characterized by frequency
$\omega$. For weak randomness  the expansion can be separated into
two distinct stages \cite{san, sha, min}: a rapid ballistic
"explosion", during the time of order $(1/\omega)$, followed by an
essentially linear evolution.   The first stage is dominated by
the nonlinearity. At the second stage, however, most of   the
interaction energy had been already converted into the kinetic
(flow) energy, so that the nonlinearity becomes weak and is
neglected \cite{foot2}. Thus, below we consider the linear
equation
\begin{eqnarray}\label{G2}
 i\hbar \, \partial_{t}\pu \Psi(\vecr, t) & = & -\frac{\hbar^2}{2m}\,\triangle\Psi(\vecr, t)
 \, + \, V(\vecr)\,\Psi(\vecr, t).
\end{eqnarray}
The initial condition for this equation is supplied by the wave function $\Phi(r)$ at the end of the first stage of the expansion, i.e., at time of order $1/\omega$. Qualitatively, this wave function is of the form \cite{castin, kagan}
\begin{eqnarray}\label{in}
\Phi(r) & = & \mathcal{F}(r)\, \exp(ir^2/a_0^2),
\end{eqnarray}
where $a_0 = (\hbar/m\omega)^{1/2}$ is the oscillator size of the
trap and $\mathcal{F}(r)$ is an envelope function which decays on the
characteristic distance $R_0\gg a_0$, where $R_0$ is the initial size of
the BEC in the trap. The envelope is often approximated by a
Gaussian, $\mathcal{F}(r) = A\, \exp(-r^2/2R_0^2)$, or by the inverted parabola
$\mathcal{F}(r) = A\sqrt{1-(r/R_0)^2}$, with $A$ being the normalization constant.

The emerging linear problem, Eqs. (\ref{G2}) and (\ref{in}), is
considerably simpler than the fermionic problem treated above.
This is because the BEC is described by a single coherent wave
function which can be treated as a classical field. Within such
mean-field description, there are no density fluctuations in the
absence of disorder. The notion of density correlations, with
their inherent statistical features, becomes meaningful only in
the presence of an external random potential. Thus, the problem
becomes similar to that considered in the theory of optical
speckles, where a classical electromagnetic wave or a scalar wave
propagates and gets scattered on a random potential \cite{akk}.
The essential difference is that in the theory of optical speckles
one usually assumes a monochromatic field, whereas the BEC wave
function, Eq. (\ref{G2}), contains a broad spectrum of wave
numbers. For a monochromatic field $\psi_{\omega}\pu(\vecr)$ with
a wave number $k_{\omega}$,  the disorder-induced
intensity-intensity correlation function, $C_{\omega}(\vecr,
\vecr')  =  \overline{\left|
\psi_{\omega}^{*}(\vecr)\psi_{\omega}\pu
 (\vecr')\right|^2}$, was calculated in
\cite{sha1},
\begin{eqnarray}
 C_{\omega}(\vecr, \vecr') & = & \overline {n_{\omega}(\vecr)} \: \overline {n_{\omega}(\vecr')} \,
  (\frac{\sin(k_{\omega}\Delta r)}
  {k_{\omega}\Delta r})^2\,
e^{-\Delta r/l_{\omega}\pu}, \: (d=3), \nonumber \\
 & & \label{intensity corr} \\ 
 C_{\omega}(\vecr, \vecr')
 & = & \overline {n_{\omega}(\vecr)} \: \overline {n_{\omega}(\vecr')} \,
  J_0^2(k_{\omega}\Delta r)\,
   e^{-\Delta r/l_{\omega}\pu}, \: (d=2), \nonumber \\
   & &  \label{intensity corr1}
\end{eqnarray}
where
$\overline{n_{\omega}(\vecr)}=\overline{\left|
\psi_{\omega}\pu(\vecr)\right|^{2}}$ is the average intensity of
the wave.

The extension to the nonmonochromatic case of a BEC is quite
straightforward. Let us start with the average density of the BEC,
$\overline{n({\bf r})}$. The BEC wave function at time $t$ is
given by
\begin{eqnarray}\label{psi}
\Psi({\bf r}, t) & = &\int\! d^{d}R\, G(\vecr, \vR, t) \, \Phi(R).
\end{eqnarray}
Using (\ref{diffuson}), one obtains in the large $t$ (and large
$r$) limit \cite{min}
\begin{eqnarray}\label{n}
\overline{n({\bf r})} & = & \overline{\mid \Psi({\bf r},t)\mid^2} \nonumber \\
& = & -\frac{1}{\pi} \int\! d\varepsilon \,P_{\varepsilon}\pu(r,t)\int
\frac{d^{d}k}{(2\pi)^d}\, {\rm Im}\bar{G}(\vk,
\varepsilon)\,|\tilde{\Phi}(k)|^2  \:\:\: \quad
\end{eqnarray}
where $\tilde{\Phi}(k)$ is the Fourier transform of the initial
condition (\ref{in}). The (equal time) field-field correlation
function, $C_{\rm field}(\Delta r,t) \equiv
\overline{\Psi^{*}(\vecr,t)\Psi(\vecr',t)}$, differs from
(\ref{n}) only by an extra factor, $f(\Delta r, \varepsilon)$,
defined in (\ref{definition f-function}),
\begin{eqnarray}\label{Cfield}
C_{\rm field}(\Delta r,t) & = & -\frac{1}{\pi} \int\! d\varepsilon
\,P_{\varepsilon}\pu(r,t)\int\! \frac{d^{d}k}{(2\pi)^d}\, f(\Delta r,
\varepsilon)\nonumber \\
& & \qquad \times {\rm Im}\bar{G}(\vk, \varepsilon)\,|\tilde{\Phi}(k)|^2 \, . \quad
\end{eqnarray}
Approximating, as above, ${\rm Im}\bar{G}(\vk, \varepsilon) =
-\pi\delta(\varepsilon-\varepsilon_{\it k}\pu)$, we obtain the
final expression for the field-field correlation function,
\begin{eqnarray}\label{Cfield1}
C_{\rm field}(\Delta r,t) & = & \int\! \frac{d^{d}k}{(2\pi)^d} \, P_k\pu(r,t) \,
f(\Delta r, \varepsilon_k) \, |\tilde{\Phi}(k)|^2 \, . \qquad
\end{eqnarray}
The short-range density-density correlation function is given by
\begin{eqnarray}\label{C}
C(\Delta r,t) & \equiv & \overline {n(\vecr)n(\vecr')}-\overline
{n(\vecr)}\,\overline{n(\vecr')} - \delta\left(\vecr-\vecr'\right)\overline
{n(\vecr)} \nonumber \\
& = & \left|C_{\rm field}(\Delta r,t)\right|^{2}.
\end{eqnarray}
Equations (\ref{Cfield1}) and (\ref{C}), supplemented by the
expression (\ref{definition f-function}) for the $f$ function, provide
the general solution for the density correlations in a BEC
diffusing in a weak random potential.

\begin{figure}
 \begin{center}
   \subfigure[]{\includegraphics[scale=0.65]{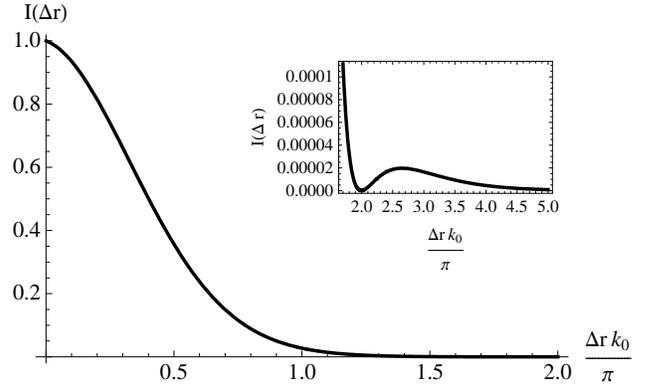} \label{fig BEC correlation function 3d}}
   \hfil
   \subfigure[]{\includegraphics[scale=0.65]{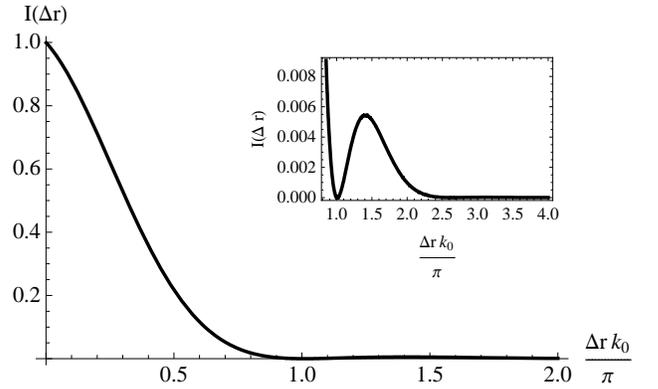} \label{fig BEC correlation function 2d}}
   \caption{The normalized density-density correlation function
$I(\Delta r,t)$ of a BEC for (a) $d=3$ and (b) $d=2$. As in
Fig. \ref{fig correlation function} the parameters were set to
$k_{0}l_{0} = 10$, $r=50 l_{0}$, $t=3r^{2}/(2d D_{k_{0}})$,
$D_{k}=D_{k_{0}} (k/k_{0})^{4-d}$,
$l_{k}=l_{0}\left(k/k_{0}\right)^{3-d}$.} \label{fig BEC
correlation function}
 \end{center}
\end{figure}

To get an estimate for the three-dimensional normalized correlation function,
\begin{eqnarray}\label{BEC normalized correlation function}
  I(\Delta r,t) & = & \frac{C(\Delta r,t)}{\overline{n(\vecr',t)}\,\overline{n(\vecr',t)}},
 \end{eqnarray}
we assume a Gaussian envelope, $\mathcal{F}(r) = A\, \exp(-r^2/2R_0^2)$. Then the Fourier transform of the BEC wave function, Eq. (\ref{in}), takes the form
\begin{eqnarray}
  |\tilde{\Phi}(k)|^{2} & = & \tilde{A}\exp(-k^2\pd/k_{0}^2),
\end{eqnarray}
with the normalization constant $\tilde{A}$ and $k_{0}\approx
2R_{0}\pu/a_{0}^2 \simeq 1/\xi$, where $\xi$ is the healing length. The
numerical evaluation of Eq. (\ref{BEC normalized correlation
function}) (see Fig. \ref{fig BEC correlation function}) shows
that, quite similar to the case of the Fermi gas, $I(\Delta r, t)$
decays on a scale $\Delta r \sim k_{0}^{-1}$. This scale remains
nearly constant in time. Since the ''cutoff function'',
$|\tilde{\Phi}(k)|^{2}$, under the integral in Eq. (\ref{Cfield1})
is not as sharp as the step function in  Eq. (\ref{correlation
function final form}), the oscillations of the normalized
correlation function are even weaker than for the fermionic case.
The short-range correlations, shown in Fig. \ref{fig BEC
correlation function}, imply that the image of a condensate,
diffusing in a weak random potential, should exhibit a random
pattern of particle density (speckle). The typical size of each
speckle spot is of the order of a few healing lengths.

\section{Conclusions}

We have studied density fluctuations in a cloud of cold atoms,
expanding in the presence of a weak random potential. Only the
simplest, two-point correlation function, $C(\Delta r,t)$, was
considered. We find that the disorder has a strong effect on
$C(\Delta r,t)$, for either a Fermi gas or for a BEC. In both
cases we obtain a random density pattern consisting of speckles of
high and low density. The typical speckle size is determined by
the decay length of the correlation function. For fermions this
size is of the order of the Fermi wavelength $\lambda_{\rm F}\pu =
2\pi/\kfermi$, where $\kfermi$ is the Fermi wave number in the
trap. It is interesting, and somewhat counterintuitive, that while
the gas keeps expanding, the typical speckle size does not change.
This is in contrast to the "clean" case of the free expansion,
when a fixed, time-independent relation exists between the
correlation length and the interparticle distance. For a BEC we
confined ourselves to a mean-field treatment, based on the
time-dependent Gross-Pitaevskii equation. In this case the speckle
structure is caused  solely by the  random potential. The typical
speckle size is given by the healing length of the condensate in
the trap and, once again, this size does not change in the process
of the expansion. If one were to go beyond the mean-field
description of a BEC, then density correlations (bunching) would
appear even in the absence of disorder. The combined effect of
quantum fluctuations and disorder on density correlations for
bosons (similarly to what has been done for fermions in Sec.
\ref{section fermi gas}) is an interesting problem which is, however, beyond the scope
of this paper \cite{foot3}. The only important parameter in our
theory is $\kfermi\lfermi$, for fermions, and $k_{0}\pu l_{0}\pu$
for the BEC. We have assumed zero temperature in all of the
calculations. The extension to finite temperatures, at least for
the Fermi case, is quite straightforward but has not been done in
the present paper.

Only short-range correlations were considered. It is well known in
optics, as well as in the mesoscopic physics of disordered
conductors, that in addition to the strong short-range
correlations there are also weak long-range correlations. Such
correlations manifest themselves as fluctuations in the
transmission coefficient through a disordered slab or as the
universal conductance fluctuations in disordered conductors
\cite{akk}. In order to observe similar effects for cold atoms one
should create a random potential in some region (say, of a shape
of a slab) and then let an atomic cloud impinge on that region.
One could then take images of the transmitted, as well as of the
reflected clouds.

Our numerical estimates have been made for a white noise random
potential, which is the case when the correlation radius of the
potential, $R_c$, is smaller than the relevant wavelength,
$\lambda$, of the matter waves (the Fermi wavelength for fermions
or the healing length for the BEC). Since the random potential for
atoms is commonly produced by creating a random pattern of light
intensity (an optical speckle), one can assume that $R_c$ is of
the order of a few light wavelengths. If $R_c$ exceeds $\lambda$,
one enters the regime of the correlated potential (colored noise).
In that regime the diffusion coefficient is determined by the
transport mean free path, $l_{\rm tr}\pu$, which is much longer than the
scattering mean free path $l_{k}\pu$ appearing in the average Green's
function. However, since both only weakly depend on $k$ our results
should remain qualitatively correct, at least in the well
developed diffusion regime, i.e., when the cloud has spread to a
distance much larger than $l_{\rm tr}\pu$. 

\acknowledgments

One of the authors (B.S.) acknowledges many useful discussions with A. Minguzzi, S. Skipetrov and B. van Tiggelen,
especially on the validity (and limitations) of the diffusion approximation. He is also indebted to J. Steinhauer for a discussion on some experimentally relevant issues. One of the authors (P.H.) acknowledges the hospitality of Technion, Haifa, where the present work was initiated. The research was supported by a grant from the Israel Science Foundation, by the Deutsche Forschungsgemeinschaft through SFB 608 and by Deutscher Akademischer Austausch Dienst (DAAD).


\begin{thebibliography}{99}
\bibitem{foot1} For a harmonic trap the effective trap size is determined by $m\omega^2L^2 \simeq \efermi$, which gives $L\simeq N^{1/2d}\sqrt{\hbar/m\omega}$.
\end{thebibliography}
\end{document}